\documentclass[useAMS,usenatbib]{mn2e} \usepackage{amsmath,amssymb}
\usepackage{graphicx,mathptm} \usepackage{times} \usepackage{natbib}
\usepackage{psfig}
\newcommand{\Mpch}{$h^{-1}\,\mbox{Mpc}$\,}
\newcommand{\hMpc}{$h\,\mbox{Mpc}^{-1}$\,}
\newcommand{\xii}{$\xi(r_\perp,r_\parallel)$\,}

\title[Neutrino effects on LSS] {Effects of Massive Neutrinos on the Large-Scale Structure of the Universe}

\author[Marulli et al.]  {Federico Marulli$^1$, Carmelita
  Carbone$^{1,4}$, Matteo Viel$^{2,3}$, Lauro
  Moscardini$^{1,4,5}$\newauthor and Andrea
  Cimatti$^1$\\ $^1$Dipartimento di Astronomia, Alma Mater Studiorum -
  Universit\`a di Bologna, via Ranzani 1, I-40127 Bologna,
  Italy\\ $^2$INAF - Osservatorio Astronomico di Trieste, Via
  G.B. Tiepolo 11, I-34131 Trieste, Italy\\ $^3$INFN/National
  Institute for Nuclear Physics, Sezione di Trieste, Via Valerio 2,
  I-34127 Trieste, Italy\\ $^{4}$INAF-Osservatorio Astronomico di
  Bologna, Via Ranzani 1, 40127, Bologna, Italy\\ $^5$INFN/National
  Institute for Nuclear Physics, Sezione di Bologna, viale Berti
  Pichat 6/2, I-40127 Bologna, Italy\\
}

\begin{document}

\maketitle

\begin{abstract}
  Cosmological neutrinos strongly affect the evolution of the largest
  structures in the Universe, i.e.  galaxies and galaxy clusters. We
  use large box-size full hydrodynamic simulations to investigate the
  non-linear effects that massive neutrinos have on the spatial
  properties of cold dark matter (CDM) haloes. We quantify the
  difference with respect to the concordance $\Lambda$CDM model of the
  halo mass function and of the halo two-point correlation function.
  We model the redshift-space distortions and compute the errors on
  the linear distortion parameter $\beta$ introduced if cosmological
  neutrinos are assumed to be massless. We find that, if not taken
  correctly into account and depending on the total neutrino mass
  $M_\nu$, these effects could lead to a potentially fake signature of
  modified gravity.  Future nearly all-sky spectroscopic galaxy
  surveys will be able to constrain the neutrino mass if $M_\nu
  \gtrsim$ 0.6 eV, using $\beta$ measurements alone and independently
  of the value of the matter power spectrum normalisation
  $\sigma_8$. In combination with other cosmological probes, this will
  strengthen neutrino mass constraints and help breaking parameter
  degeneracies.
\end{abstract}

\begin{keywords} 
  cosmology: theory -- cosmology: observations, dark matter,
  neutrinos, galaxy clustering
\end{keywords}

\section {Introduction}
\label{intro}

Neutrinos are so far the only dark matter candidates that we actually
know to exist. Since deviations from the standard model in the form of
extra neutrino species are still uncertain \citep{giusarma2011,
  gonzalez-morales2011}, in this paper we focus on standard neutrino
families only.  It is now established from solar, atmospheric, reactor
and accelerator neutrino experiments that neutrinos have non-zero mass
implying a lower limit on the total neutrino mass given by
$M_\nu\equiv \sum m_{\nu}\sim 0.05$ eV \citep{lesgourgues2006}, where
$m_\nu$ is the mass of a single neutrino species. On the other hand
the absolute masses are still unknown. Since neutrino mass affects the
evolution of the Universe in several observable ways, its measurements
can be obtained from different cosmological probes as observations of
Cosmic Microwave Background (CMB), galaxy clustering, Ly$\alpha$
forest, and weak lensing data \citep{abazajian2011}.

In particular, a thermal neutrino relic component in the Universe
impacts both the expansion history and the growth of cosmic
structures. Neutrinos with mass $\lesssim 0.6$ eV become
non-relativistic after the epoch of recombination probed by the CMB,
and this mechanism allows massive neutrinos to alter the
matter-radiation equality for a fixed $\Omega_m h^2$
\citep{lesgourgues2006}. Massive neutrinos act as non-relativistic
particles on scales $k>k_{\rm nr}=0.018(m_\nu/1{\rm
  eV})^{1/2}\Omega_m^{1/2}$ \hMpc, where $k_{\rm nr}$ is the
wave-number corresponding to the Hubble horizon size at the epoch
$z_{\rm nr}$ when the given neutrino species becomes non-relativistic,
$\Omega_m$ is the matter energy density and $h=H_0/100\, {\rm km\,
  s^{-1} Mpc^{-1}}$. The large velocity dispersion of non-relativistic
neutrinos suppresses the formation of neutrino perturbations in a way
that depends on $m_\nu$ and redshift $z$, leaving an imprint on the
matter power spectrum for scales $k>k_{\rm fs}(z)=0.82
H(z)/H_0/(1+z)^2 (m_\nu/1{\rm eV})$ \hMpc
\citep{takada2006,lesgourgues2006}, where neutrinos cannot cluster and
do not contribute to the gravitational potential wells produced by
cold dark matter and baryons.  This modifies the shape of the matter
power spectrum and the correlation function on these scales \citep[see
  e.g.][and reference
  therein]{doroshkevich1981,hu1998,abazajian2005,kiakotou2008,brandbyge2010,viel2010}.

Massive neutrinos affect also the CMB statistics. WMAP7 alone
constrains $M_\nu < 1.3$ eV \citep{komatsu2010} and, thanks to the
improved sensitivity to polarisation and to the angular power spectrum
damping tail, forecasts for the Planck satellite alone give a
1--$\sigma$ error on the total neutrino mass of $\sim 0.2-0.4$ eV,
depending on the assumed cosmological model and fiducial neutrino mass
\citep[e.g.][and references therein]{perotto2006, kitching2008}.
Moreover, the combination of present data-sets from CMB and
large-scale structure (LSS) yields an upper limit of $M_\nu<0.3$ eV
\citep[e.g.]{wang2005,vikhlinin2009,thomas2010,gonzalez2010,reid2010}.
A further robust constraint on neutrino masses has been obtained using
the Sloan Digital Sky Survey flux power spectrum alone, finding an
upper limit of $M_\nu<0.9$ eV ($2\sigma$ C. L.)
\citep{viel2010}. However, the tightest contraints to date in terms of
a 2$\sigma$ upper limit on the neutrino masses have been obtained by
combining the Sloan Digital Sky Survey flux power from the Lyman alpha
forest with CMB and galaxy clustering data and result in $\Sigma
m_\nu<0.17$ eV \citep{seljak2006}. Somewhat less constraining bounds
have been obtained by \citet{goobar2006}, while for forecasting on
future joint CMB and Lyman-alpha constraints we refer to
\citet{gratton2008}.  For further discussion on neutrino mass
constraints from different probes see e.g. \citet{abazajian2011} and
reference therein.

The forecasted sensitivity of future LSS experiments, when combined
with Planck CMB priors, indicates that observations should soon be
able to detect signatures of the cosmic neutrino background and
measure the neutrino mass even in the case of the minimum mass
$M_\nu=0.05$ eV
\citep[e.g.][]{hannestad2007,kitching2008,lsst2009,hannestad2010,lahav2010}.
In particular, \citet{carbone2011} show that future spectroscopic
galaxy surveys, such as EUCLID, JEDI and WFIRST, not only will be able
to measure the dark-energy equation of state with high accuracy, but
they will determine the neutrino mass scale independently of flatness
assumptions and dark energy parametrization, if the total neutrino
mass $M_\nu$ is $>0.1$ eV. On the other hand, if $M_\nu$ is $<0.1$ eV,
the sum of neutrino masses, and in particular the minimum neutrino
mass required by neutrino oscillations, can be measured in the context
of a $\Lambda$CDM model.

It is therefore mandatory to measure with high accuracy the growth
history of large-scale structures in order to obtain the necessary
cosmological information, excluding possible systematics due to the
incorrect assumption that neutrinos are massless.  One way of
determining the growth of structure is through the redshift-space
distortions (RSD) of the galaxy distribution, caused by the
line-of-sight component of galaxy peculiar velocities.  RSD can
be exploited in large deep redshift surveys to measure (if the galaxy
bias is measured independently) the growth rate of density
fluctuations $f\equiv d\ln D/d\ln a$, with $D$ being the linear
density growth factor and $a=1/(1+z)$, or to measure the linear
redshift-space distortion parameter $\beta$ \citep{kaiser1987} that
depends on the growth rate $f$ and the galaxy linear bias $b$.  In
particular, $\beta$ in the presence of massive neutrinos depends on
both redshift and wave-numbers $\beta(z,k)=f(z,k)/b(z)$, since in this
case the linear growth rate $f(z,k)$, being suppressed by
free-streaming neutrinos, acquires a \emph{scale dependence} already
at the linear level \citep{kiakotou2008}.

There are two types of RSD with competing effects acting along
opposite directions on the observed galaxy correlation function.
While, for large separations, large-scale bulk peculiar velocities
produce a flattening effect on the correlation function and give
information on the growth of structures, on small scales, random
peculiar velocities cause the so-called Fingers of God (FoG),
stretching compact structures along the line-of-sight
\citep{scoccimarro2004,song2009}.

RSD have been the subject of many analyses, as reviewed in
\citet{hamilton1998}. The latest large galaxy surveys that have
enabled measurements of RSD via the correlation function and the power
spectrum are the 2-degree Field Galaxy Redshift Survey
\citep{peacock2001,hawkins2003,percival2004} and the Sloan Digital Sky
Survey
\citep{tegmark2004,zehavi2005,tegmark2006,okumura2008,cabre2009,cabre2009b}.
Moreover, also the VIMOS-VLT Deep Survey have been exploited in
\citet{guzzo2008} for RSD determinations from the correlation
function.

Since the linear theory description is valid only at very large
scales, an extension of the theoretical description has been attempted
to quasi-linear and non-linear scales using empirical methods based on
the so-called streaming model \citep{peebles1980}, consisting of
linear theory and a convolution on the line-of-sight with a velocity
distribution. This model describes the FoG elongation along the
line-of-sight due to random motions of virialised objects
\citep{jackson1972}. It has been shown by \citet{guzzo2008},
\citet{cabre2009} and \citet{percival2009} that on quasi-linear scales
a streaming model with a Gaussian velocity dispersion is a good
general fit to the redshift-space power spectrum. However, this model
is not accurate on very small and very large scales
\citep{taruya2010,okumura2011,raccanelli2011} and fitting functions
based on simulation results have been used
\citep{hatton1999,scoccimarro2004,tinker2006,tinker2007,tocchini2011}.
Anyway, to the purpose of this paper, the streaming model is accurate
enough to robustly constrain the effect of massive neutrinos on RSD
when applied on scales $\lesssim 50$ \Mpch.

In this work, we compare analytic results against a set of large
N-body hydrodynamical simulations developed with an extended version of
{\small GADGET III}, that is an improved version of the code described in
\citet{springel2005gadget2}, further modified to take into account the
effect of massive free-streaming neutrinos on the evolution of cosmic
structures \citep{viel2010}.  It is well known that
galaxy/halo bias on large, linear scales is scale-independent, but
becomes non-linear and therefore scale-dependent on smaller
scales. This effect can be mimicked or enhanced by the presence of
massive neutrinos.  Therefore, the effect of massive neutrinos on the
galaxy clustering in the quasi non-linear regime has to be explored
via N-body simulations to encompass all the relevant effects, and
analyse possible sources of systematic errors due to non-linearities
and galaxy bias scale-dependence.  In particular, in this work we will
focus on the DM halo mass function (MF), the DM halo bias and RSD.

The rest of the paper is organised as follows. In \S \ref{formalism} we
review our method and the adopted modelling of RSD. In \S \ref{N-body} we
describe the exploited set of N-body simulations and present our
results on the neutrino effects on LSS. Finally in \S \ref{conclu}
we draw our conclusions.

\section {Formalism to model redshift distortions}
\label{formalism}
\subsection {Overview}
\label{RSD}

In this section we describe how RDS are generated in the observed
galaxy correlation function.  An observed galaxy redshift is composed
by the two additive terms,
\begin{equation}
z_{\rm obs}=z_{\rm c}+\frac{v_\parallel}{c}(1+z_{\rm c}),
\label{eq:redshift}
\end{equation}
where $z_{\rm c}$ is the {\em cosmological} redshift, due to the
Hubble flow. The second term of Eq.~(\ref{eq:redshift}) is caused by galaxy
peculiar velocities where $v_\parallel$ is the component parallel to
the line-of-sight. The {\em real} comoving distance of a galaxy is
given by
\begin{equation}
r_\parallel=c\int_0^{z_{\rm c}}\frac{dz'_{\rm c}}{H(z'_{\rm c})},
\label{eq:distance}
\end{equation}
where $H(z'_{\rm c})$ is the so-called Hubble rate.  When the distances are
computed replacing $z_{\rm c}$ with $z_{\rm obs}$ in
Eq.~(\ref{eq:distance}), 
i.e. without correcting for the peculiar
velocity contribution, we say to be in the {\em redshift-space}. We
will refer to the redshift-space spatial coordinates using the vector
$\vec{s}$, while we will use $\vec{r}$ to indicate the real-space
coordinates.

Fundamental information is hinted in the anisotropies of an observed
galaxy map in redshift-space.  A useful statistics widely used to
describe the spatial properties of a general astronomical population
is the two-point correlation function, $\xi(r)$, implicitly defined as
$dP_{12} = n^2[1+\xi(r)]dV_1dV_2$, where $dP_{12}$ is the probability
of finding a pair with one object in the volume $dV_1$ and the other
in the volume $dV_2$, separated by a comoving distance $r$. It is
convenient to decompose the distances into the two components
perpendicular and parallel to the line-of-sight,
$\vec{r}=(r_\perp,r_\parallel)$, so that the correlation becomes a
two-dimensional function of these variables.  When measured in
real-space, the contour lines of \xii are circles for an isotropic
population of objects as galaxies. Instead, in redshift-space $\xi$ is
distorted: at small scales ($\lesssim$1 \Mpch) the distortion is caused
by the random motions of galaxies moving inside virialised
structures. This motion changes the shape of $\xi$ in the direction
parallel to the line-of-sight, producing the observed FoG. At large
scales the coherent bulk motion of virialising structures squashes the
correlation function $\xi$ perpendicularly to the line-of-sight.

A different kind of distortion, called {\em geometrical} or
Alcock-Paczynski (AP) distortion \citep{alcock1979}, can be present if
the cosmological parameters assumed in Eq.~(\ref{eq:distance}) are not
the same as the true cosmological model of the Universe. In what
follows, we will not consider this effect since massive neutrinos do
not produce a geometrical distortion if the present-day total matter
energy density parameter $\Omega_m$ is held fixed.
\begin{figure*}
\includegraphics[width=0.48\textwidth]{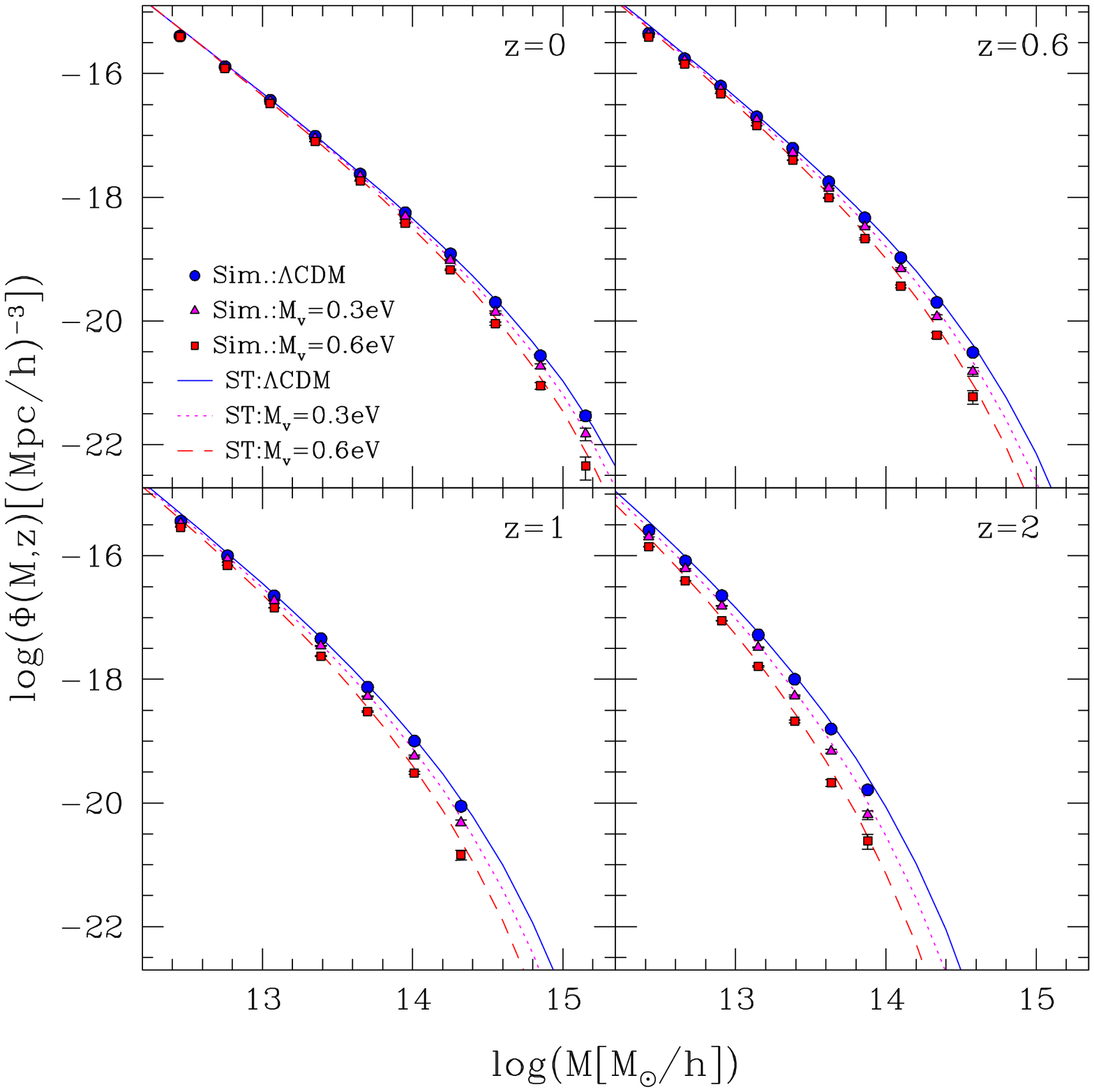}
\includegraphics[width=0.48\textwidth]{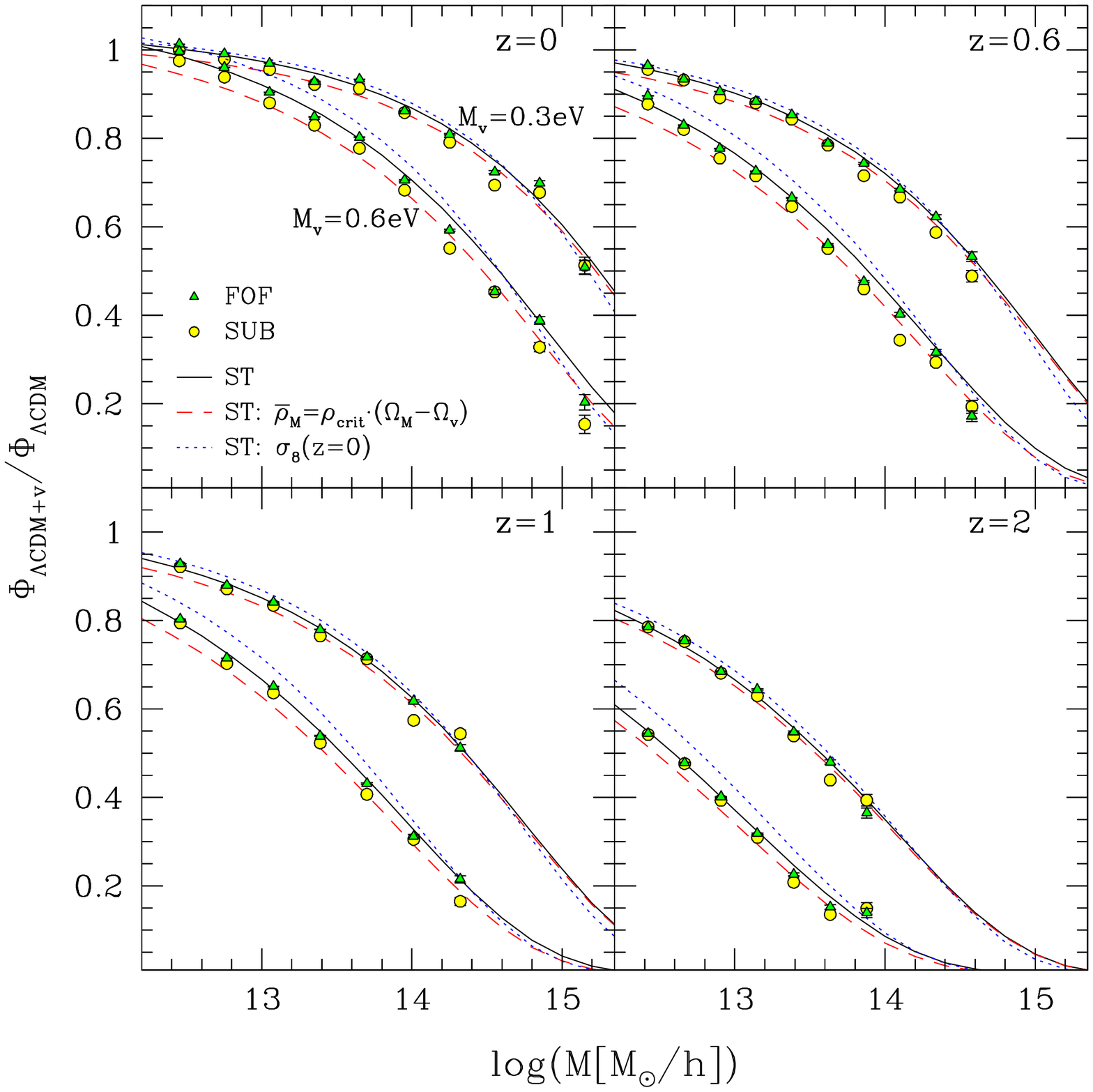}
\caption{DM halo MF as a function of $M_\nu$ and redshift. Left: MF of
  the SUBFIND haloes in the $\Lambda$CDM N-body simulation (blue
  circles) and in the two simulations with $M_\nu=0.3$ eV (magenta
  triangles) and $M_\nu=0.6$ eV (red squares). The blue, magenta and
  red lines show the halo MF predicted by \citet{sheth2002}, where the
  variance in the density fluctuation field, $\sigma(M)$, for the
  three cases, $M_\nu=0,0.3,0.6$ eV, has been computed using the
  linear matter $P(k)$ extracted from {\small CAMB}. Right: ratio
  between the halo MFs of the simulations with and without
  neutrinos. The green triangles show the MF ratios of the FoF haloes,
  while yellow circles show the ones of the SUBFIND haloes.  The lines
  represent the ST-MF ratios: the black solid lines are the MF ratios
  predicted for $M_\nu=0.3$ eV and $M_\nu=0.6$ eV; the red dashed
  lines are the same ratios but assuming
  $\bar{\rho}=\rho_{c}\cdot(\Omega_m-\Omega_\nu)$ in the ST-MF formula
  Eq.~(\ref{eq:MF}) \citep{brandbyge2010}; finally, the blue dotted
  lines are the ratios between the ST MFs in two $\Lambda$CDM
  cosmologies, which differ for the $\sigma_8$ normalisation, as
  explained in the text.  The error bars represent the statistical
  Poisson noise.}
 \label{fig:MF}
\end{figure*}

\subsection{Modelling the dynamical distortions}
\label{modelling}

At large scales and in the plane-parallel approximation, the dynamical
distortions can be parameterised in the Fourier space as follows
\begin{equation}
  P(k) = (1+\beta\mu^2)^2 P_{\rm lin}(k),
\label{eq:kaiser}
\end{equation}
where $P_{\rm lin}(k)$ is the linear power spectrum of the matter
density fluctuations and $\mu$ is the cosine of the angle between
$\vec{k}$ and the line-of-sight.
Fourier transforming equation (\ref{eq:kaiser}) gives
\begin{equation} 
\xi(s,\mu)= \xi_0(s)P_0(\mu)+\xi_2(s)P_2(\mu)+\xi_4(s)P_4(\mu),
\label{eq:ximodellin}
\end{equation}
where the functions $P_l$ represent the Legendre polynomials
\citep{hamilton1992}. The multipoles $\xi_n(s)$, $n=0,2,4$, can be written as
follows
\begin{equation} 
\xi_0(s) = \left(1+ \frac{2\beta}{3} + \frac{\beta^2}{5}\right)\xi(r),
\end{equation}
\begin{equation} 
\xi_2(s) = \left(\frac{4\beta}{3} +
\frac{4\beta^2}{7}\right)[\xi(r)-\overline{\xi}(r)],
\end{equation}
\begin{equation} 
\xi_4(s) = \frac{8\beta^2}{35}\left[\xi(r) +
  \frac{5}{2}\overline{\xi}(r)
  -\frac{7}{2}\overline{\overline{\xi}}(r)\right],
\end{equation}
where $\xi(r)$ is the real-space correlation function, and the {\em
  barred} correlation functions are defined as
\begin{equation} 
\overline{\xi}(r) = \frac{3}{r^3}\int^r_0dr'\xi(r')r'{^2},
\end{equation}
\begin{equation}
\overline{\overline{\xi}}(r) = \frac{5}{r^5}\int^r_0dr'\xi(r')r'{^4}.
\end{equation}
Eq.~(\ref{eq:ximodellin}) can be used to approximate the correlation
function at large scales. To include in the model also the small
scales, as discussed in \S \ref{intro}, we adopt the streaming model
and use the following formula
\begin{equation} 
 \xi(s_\perp, s_\parallel) = \int^{\infty}_{-\infty}dv
 f(v)\xi(s_\perp, s_\parallel - v/H(z)/a(z)),
\label{eq:ximodel}
\end{equation}
where $f(v)$ is the distribution function of random pairwise
velocities that are measured in physical (not comoving) coordinates
\citep[but see e.g.][]{scoccimarro2004, matsubara2004}. In this work, we
adopt for $f(v)$ the form
\begin{equation}
f(v)=\frac{1}{\sigma_{12}\sqrt{2}}
\exp\left(-\frac{\sqrt{2}|v|}{\sigma_{12}}\right),
\label{eq:fvexp} 
\end{equation}
where $\sigma_{12}$ is the dispersion in the pairwise peculiar
velocities.

\section{The adopted set of numerical simulations}
\label{N-body}

The set of simulations we consider in this work have been performed by
\citet{viel2010} with the hydrodynamical TreePM-SPH (Tree Particle
Mesh-Smoothed Particle Hydrodynamics) code {\small GADGET III}, which is
an improved and extended version of the code described in
\citet{springel2005gadget2}.  This code has been modified in order to
simulate the evolution of the neutrino density distribution.  The
cosmological model adopted in the simulations is based on cold dark
matter and assumes the presence of the cosmological constant
($\Lambda$CDM): $n_s=1$, $\Omega_m=0.3$, $\Omega_b=0.05$,
$\Omega_{\Lambda}=0.7$ and $h=0.7$ ($H_0=100\,h$ km/s), plus a
cosmological massive neutrino component $\Omega_\nu\equiv M_\nu/(h^2
93.8 {\rm eV})$ ($\Lambda$CDM+$\nu$).  In what follows, we consider
only the so-called ``grid based implementation'' of the simulations
developed by \citet{viel2010}, where neutrinos are treated as a fluid
\citep[see also][]{brandbyge2009, brandbyge2010b}.  In this
implementation the linear growth of the perturbations in the neutrino
component is followed by interfacing the hydrodynamical code with the
public available Boltzmann code {\small
  CAMB}\footnote{http://camb.info/} \citep{lewis2000}.  More
specifically, the power spectra of the neutrino density component are
interpolated in a table produced via {\small CAMB} of one hundred
redshifts in total, spanning logarithmically the range $z=0-49$. The
gravitational potential is calculated at the mesh points and the
neutrino contribution is added when forces are calculated by
differentiating this potential.

In this approach the gravitational force due to neutrinos is
calculated based on the linearly evolved density distribution of the
neutrinos in Fourier space. This implementation has the advantage that
it does not suffer from significant shot noise on small scales,
yielding therefore higher accuracy at scales and redshifts where the
effect of the non-linear neutrino evolution is still moderate,
especially for small neutrino masses.  Further advantages of such a
grid based approach, aside from eliminating the Poisson noise, are the
reduced requirements with regard to memory (there are no neutrino
positions and velocities to be stored) and computational time.

The initial conditions of this set of simulations were generated based
on linear matter power spectra separately computed for each component
(dark matter, gas and neutrinos) with {\small CAMB}.  The total matter
power spectrum was normalised such that its amplitude (expressed in
terms of $\sigma_8$) matches the {\small CAMB} prediction at the
same redshift.  The mass per simulation particle at our default
resolution is $1.4\cdot10^{10}M_\odot/h$ and $6.9\cdot 10^{10}M_\odot/h$ for
gas and dark matter, respectively.  In this work we
consider the set of simulations with a box of comoving volume $V=(512\,
h^{-1}\,{\rm Mpc})^3$, and total neutrino mass $M_\nu=0, 0.3, 0.6$ eV,
respectively.

To identify DM haloes and their substructures we have used two
different algorithms: a standard Friends-of-Friends (FoF) group-finder
with linking length $b=0.2$, and the {\small SUBFIND} algorithm
described in \citet{springel2001b}. Apart from the right panel of
Fig.~\ref{fig:MF}, all the results presented in this paper have been
obtained using our sub-halo catalogues, composed by the
gravitationally bound substructures that SUBFIND identifies in each
FoF halos. However, as we have explicitly verified, all the main
conclusions of this work do not change if we consider the halo
catalogues instead.  This happens because, as showed by
\citet{giocoli2010}, in the mass range considered in this work the
total mass function of haloes and sub-haloes is manly dominated by the
halo systems. At $z=0$ the sub-halo contribution start to be seen only
for masses $\lesssim 10^{10} M\odot/h$.  For all the considered
$M_\nu$ values, we have restricted our analysis in the mass range
$M_{\rm min}<M<M_{\rm max}$, where $M_{\rm min}=2\cdot10^{12} M_\odot/h$ and $M_{\rm
  max}=2\cdot10^{15},5\cdot10^{14},3\cdot10^{14},10^{14} M_\odot/h$ at
$z=0,0.66,1,2$, respectively.

\section {Results}
\label{Results} 

In this Section we show how the halo MF, the halo bias, and the linear
redshift-space distortion parameter $\beta$ get modified with respect
to the standard $\Lambda$CDM case, when a massive neutrino component
is taken correctly into account.  In particular we compare the results
between the $\Lambda$CDM and the $\Lambda$CDM+$\nu$ cosmologies, and
analyse if our findings agree with analytical predictions in the
literature.

\begin{figure*}
\includegraphics[width=\textwidth]{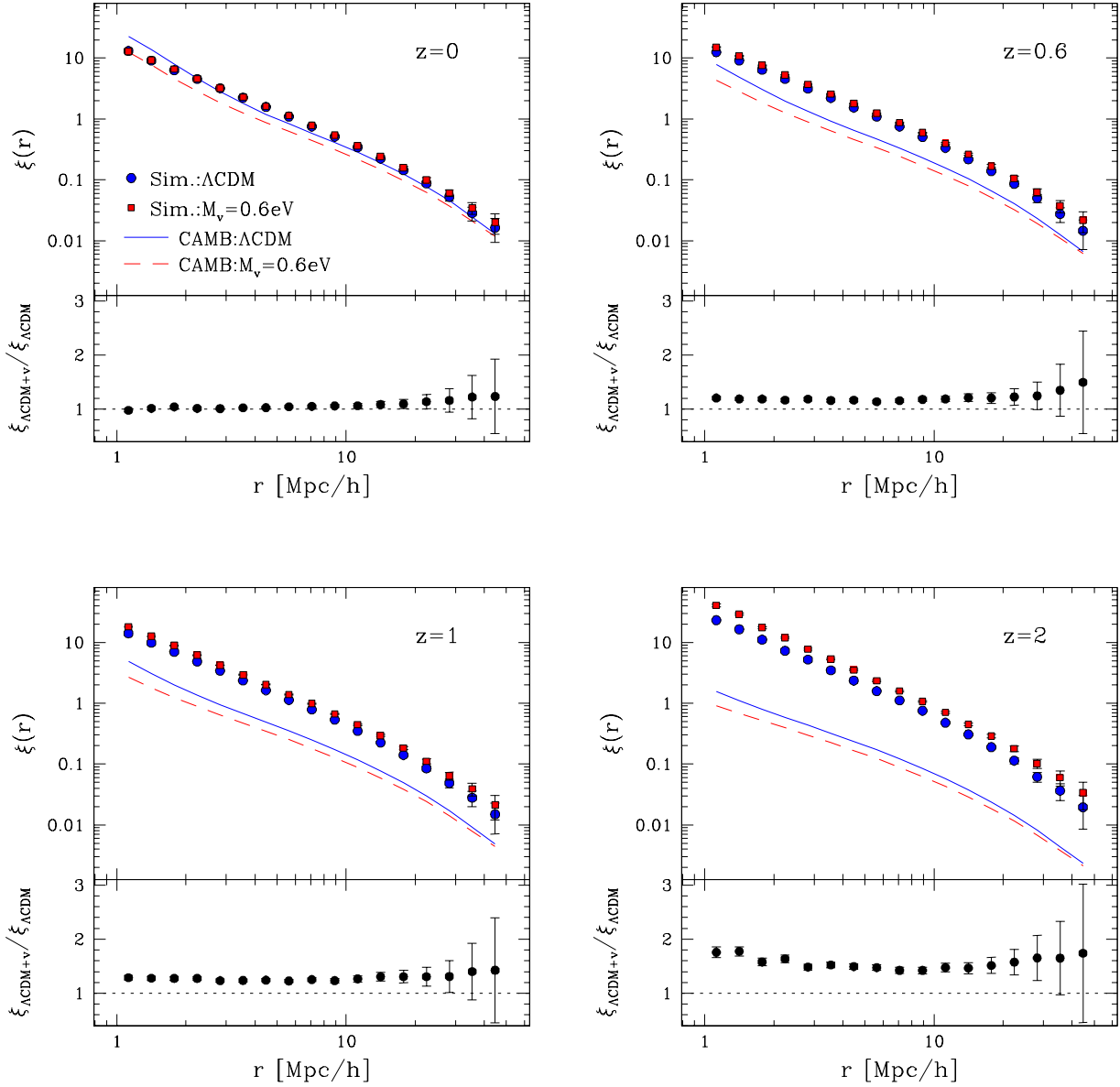}
\caption{Real-space two-point auto-correlation function of the DM
  haloes in the $\Lambda$CDM N-body simulation (blue circles) and in
  the simulation with $M_\nu=0.6$ eV (red squares). The blue and red
  lines show the DM correlation function, for $M_\nu=0$ and
  $M_\nu=0.6$ eV, respectively, obtained by Fourier trasforming the
  non-linear power spectrum extracted from CAMB \citep{lewis2002}
  which exploits the HALOFIT routine \citep{smith2003}. The bottom
  panels show the ratio between the halo correlation function of the
  simulations with and without neutrinos. The error bars represent the
  statistical Poisson noise corrected at large scales as prescribed by
  \citet{mo1992}.}
 \label{fig:xiReal}
\end{figure*}

\subsection{The halo mass function}
\label{MF}

As mentioned in \S \ref{intro}, the free-streaming of non-relativistic
neutrinos contrasts the gravitational collapse which is the basis of
cosmic structure formation. The first consequence of this mechanism is
represented by a significant suppression in the average number density
of massive structures. This effect can be observed in the high mass
tail of the halo MF as measured from our set of simulations, and shown
by the data points in the left panel of Fig.~\ref{fig:MF}. For a fixed
amplitude of the primordial curvature perturbations $\Delta^2_{\cal
  R}$, the amount of the number density suppression depends on the
value of the total neutrino mass $M_\nu$. From the comparison of the
corresponding MFs, we recover what theoretically expected, i.e.  the
higher the neutrino mass is, the larger the suppression in the
comoving number density of DM haloes becomes. The suppression affects
mainly haloes of mass $10^{14} M_\odot/h <M<10^{15} M_\odot/h$,
depending slightly on the redshift $z$. This result is in agreement
with the findings of \citet{brandbyge2010}.  In the same plot, we
compare the measured MFs with the analytical predictions of
\citet{sheth2002} (ST), represented by the solid, dotted and dashed
curves, corresponding to the values $M_\nu=0,0.3,0.6$ eV,
respectively.  The ST fit is based on the fact that the halo MF can be
written as \citep{press1974}
\begin{equation}
\frac{M dM}{\bar \rho}\frac{dn(M,z)}{dM} = \zeta f(\zeta)
\frac{d\zeta}{\zeta},
\label{eq:MF}
\end{equation}
with $\zeta \equiv [\delta_{\rm sc}(z)/\sigma(M)]^2$, where
$\delta_{\rm sc}(z)=1.686$ is the overdensity required for spherical
collapse at $z$ in a $\Lambda$CDM cosmology, and $\bar \rho = \Omega_m
\rho_c$, where $\rho_c$ is the critical density of the Universe.  Here
$\Omega_m =\Omega_{cdm}+\Omega_b+\Omega_\nu$, and $dn(M,z)$ is the
number density of haloes in the mass interval $M$ to $M + dM$.  The
variance of the linear density field, $\sigma^2(M)$, is given by
\begin{equation}
\sigma^2(M) = \int dk \frac{k^2 P_{\rm lin}(k)}{2 \pi^2}
|W(kR)|^2,
\end{equation}
where the top-hat window function is $W(x) = (3/x^3)(\sin x - x \cos
x)$, with $R = (3M/4\pi \bar \rho)^{1/3}$.

The ST fit to $\zeta f(\zeta)$ is
\begin{equation}
\zeta f(\zeta) = A
\left(1+\frac{1}{\zeta'^p}\right)\left(\frac{\zeta'}{2}\right)^{1/2}
\frac{e^{-\zeta'/2}}{\sqrt{\pi}},
\end{equation}
with $\zeta' = 0.707 \zeta$, $p=0.3$, and $A=0.3222$ determined from
the integral constraint $\int f(\zeta) d\zeta = 1$.

\begin{figure}
\includegraphics[width=0.48\textwidth]{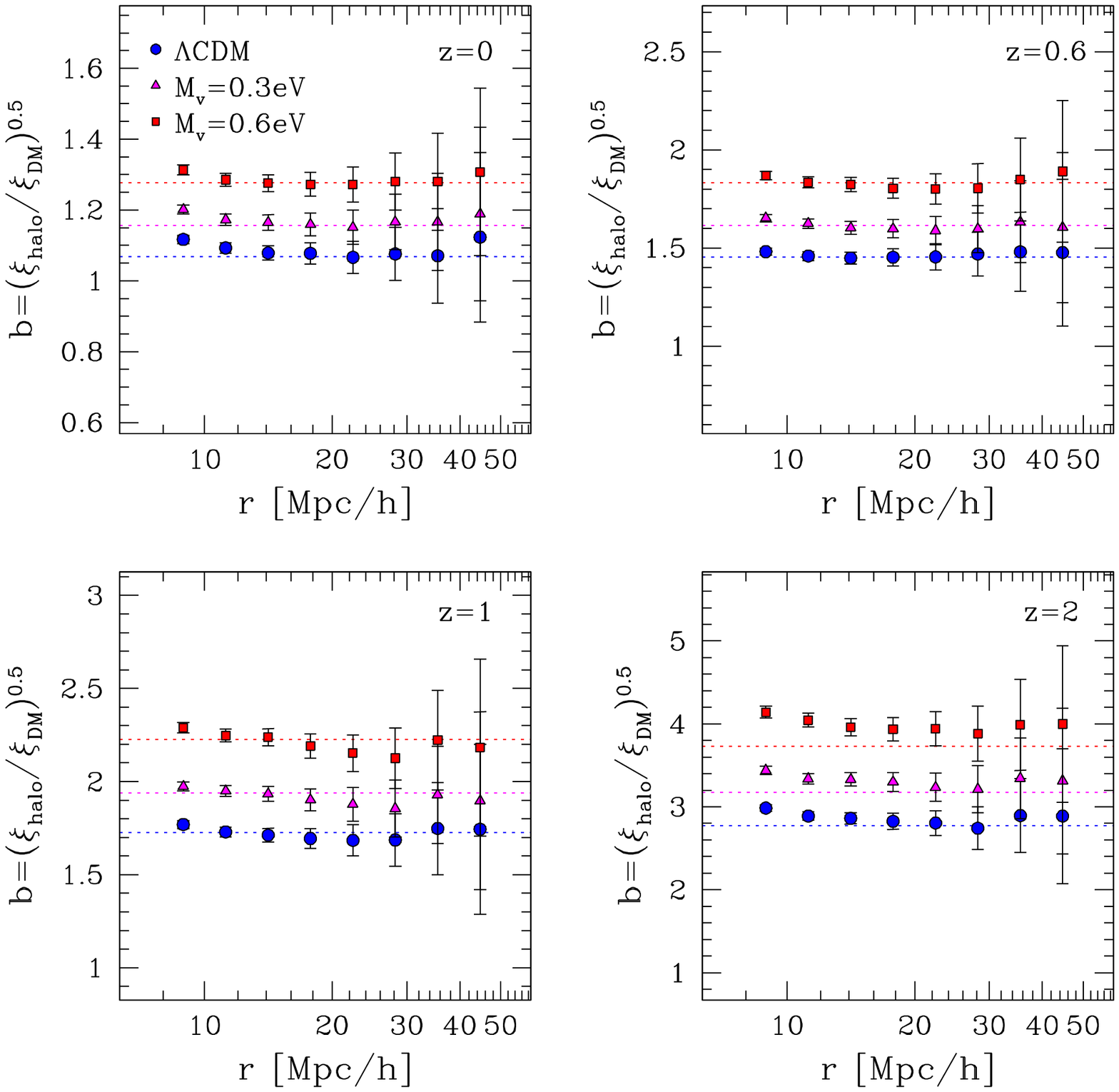}
\caption{DM halo bias, $b=(\xi_{\rm halo}/\xi_{\rm DM})^{0.5}$,
  measured from the $\Lambda$CDM simulations (blue circles) and from
  the two simulations with $M_\nu=0.3$ eV (magenta triangles) and
  $M_\nu=0.6$ eV (red squares). The error bars represent the
  propagated Poisson noise corrected at large scales as
  prescribed by \citet{mo1992}. Dotted lines are the theoretical
  predictions of \citet{sheth_mo_tormen2001}
  (Eq.~(\ref{eq:bias})). The four panels show the results at different
  redshifts, as labeled.}
 \label{fig:bias1}
\end{figure}

In Fig.~\ref{fig:MF} the variance of the density field, $\sigma^2(M)$,
has been computed with the matter power spectrum extracted from
{\small CAMB} \citep{lewis2000}, using the same cosmological
parameters of the simulations.  In particular, in the left panel we
show the MF of sub-structures identified using the SUBFIND algorithm,
where the normalisation of the matter power spectrum is fixed by
$\Delta^2_{\cal R}(k_0)=2.3\times 10^{-9}$ at $k_0=0.002$/Mpc
\citep{larson2011}, chosen to have the same value both in the
$\Lambda$CDM+$\nu$ and in the $\Lambda$CDM cosmologies. The error bars
represent the statistical Poisson noise.

In the right panel of Fig.~\ref{fig:MF} we show the ratios between the
halo MFs evaluated in the $\Lambda$CDM+$\nu$ and $\Lambda$CDM
cosmologies. In particular, the triangles represent the ratios of the
halo MFs measured directly from the simulations after identifying the
structures via a FoF group finder \citep{springel2005gadget2}, and the
filled circles represent the ratios between the $\Lambda$CDM+$\nu$ and
$\Lambda$CDM MFs for the substructure evaluated via the SUBFIND
algorithm.

\begin{figure}
\includegraphics[width=0.48\textwidth]{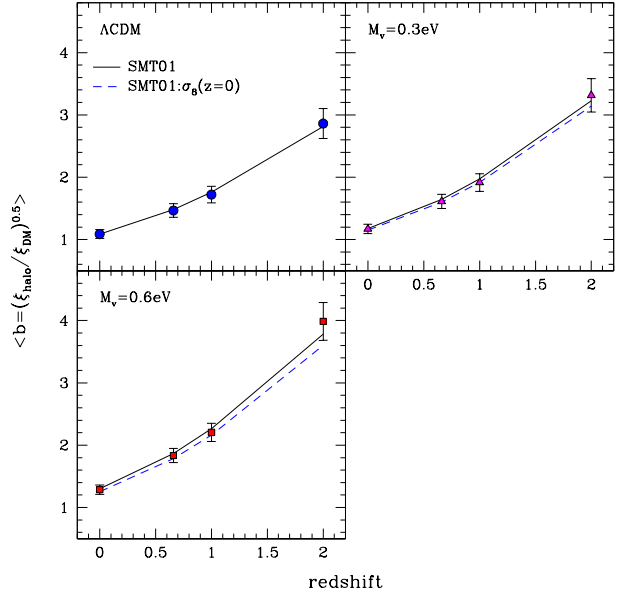}
\caption{Mean bias (averaged in
  $10\,h^{-1}\,\mbox{Mpc}<r<50\,h^{-1}\,\mbox{Mpc}$) as a function of
  redshift compared to the theoretical predictions of
  \citet{sheth_mo_tormen2001} (dotted lines)
  (Eq.~(\ref{eq:bias})). Here the dashed lines represent the
  theoretical expectations for a $\Lambda$CDM cosmology renormalized
  with the $\sigma_8$ value of the simulations with a massive neutrino
  component. The error bars represent the propagated Poisson noise
  corrected at large scales as prescribed by \citet{mo1992}.}
 \label{fig:bias2}
\end{figure}

The curves in the panel show the corresponding ST predictions for
three cases: \emph{i}) the solid lines represent the ratios between
the theoretical MFs when the total
$\Omega_m=\Omega_{cdm}+\Omega_b+\Omega_\nu$ is inserted in
Eq.~(\ref{eq:MF}) through the expression $\bar\rho=\Omega_m \rho_c$;
\emph{ii}) the dashed lines represent the theoretical MF ratios when
the quantity $\bar\rho=(\Omega_{cdm}+\Omega_b)\rho_c$ is used in
Eq.~(\ref{eq:MF}); \emph{iii}) finally, the dotted lines represent the
ratios between the ST MFs in two $\Lambda$CDM cosmologies, which
differ for the $\sigma_8$ normalisation, i.e. the ratio between the MF
in a $\Lambda$CDM cosmology having the same $\sigma_8$ value of the
simulations with a massive neutrino component, and the MF in a
$\Lambda$CDM cosmology having a $\sigma_8$ value in agreement with the
CMB normalisation $\Delta^2_{\cal R}(k_0)=2.3\times 10^{-9}$ and in
the absence of massive neutrinos.

We note that the MFs of the haloes obtained with the FoF algorithm
look to be better fitted by the theoretical predictions of the
\emph{i})-case, while the MFs of the substructures obtained with the
SUBFIND algorithm have a trend much more similar to the predictions of
the \emph{ii})-case \citep{brandbyge2010}.  Moreover, we see that, as
the redshift increases, the suppression of the halo number density due
massive neutrinos moves also towards masses $M \leq 10^{14}
M_\odot/h$. As an example, the number density of haloes with mass
$10^{14} M_\odot/h$ at $z=0$ decreases by $\sim 15\%$ for $M_\nu=0.3$ eV
and by $\sim 30\%$ for $M_\nu=0.6$ eV, and, at $z=1$, by$\sim 40\%$
and $\sim 70\%$, respectively.

As already discussed in \citet{viel2010}, the free-straming of massive
neutrinos leads to the well known degeneracy between the values of
$\sigma_8$ and $M_\nu$. However, as the \emph{iii})-case lines show,
the difference between the MFs with and without neutrinos does not
reduce merely to a $\sigma_8$ renormalisation of the background
cosmology, since, even renormalising to the same $\sigma_8$, neutrinos
free-streaming alters the MF, changing its shape and amplitude
especially for the less massive objects (compare the dotted-blue and
solid-black lines in the right panel of Fig.~\ref{fig:MF}). However,
the possibility to measure this effect and break the
$M_\nu$-$\sigma_8$ degeneracy depends on the value of the neutrino
mass and on the sensitivity in measuring the MF at masses
$M<10^{14}M_\odot/h$. The resolution limits of our simulations do not
allow us to study the properties of haloes with masses
$M\lesssim10^{12}M_\odot/h$. Future galaxy surveys, like EUCLID,
should be able to break this degeneracy measuring the number counts of
low mass galaxies together with the overall shape of the matter power
spectrum \citep[see e.g.][]{carbone2011}. Moreover, even with present
observational data it has already been possible to break the
$M_\nu$-$\sigma_8$ degeneracy. For instance, \citet{viel2010} robustly
constrained the neutrino masses using the Sloan Digital Sky Survey
flux power spectrum alone, without CMB priors on $\sigma_8$.

\subsection{The halo clustering and bias}
\label{clustering}

\begin{figure*}
\includegraphics[width=\textwidth]{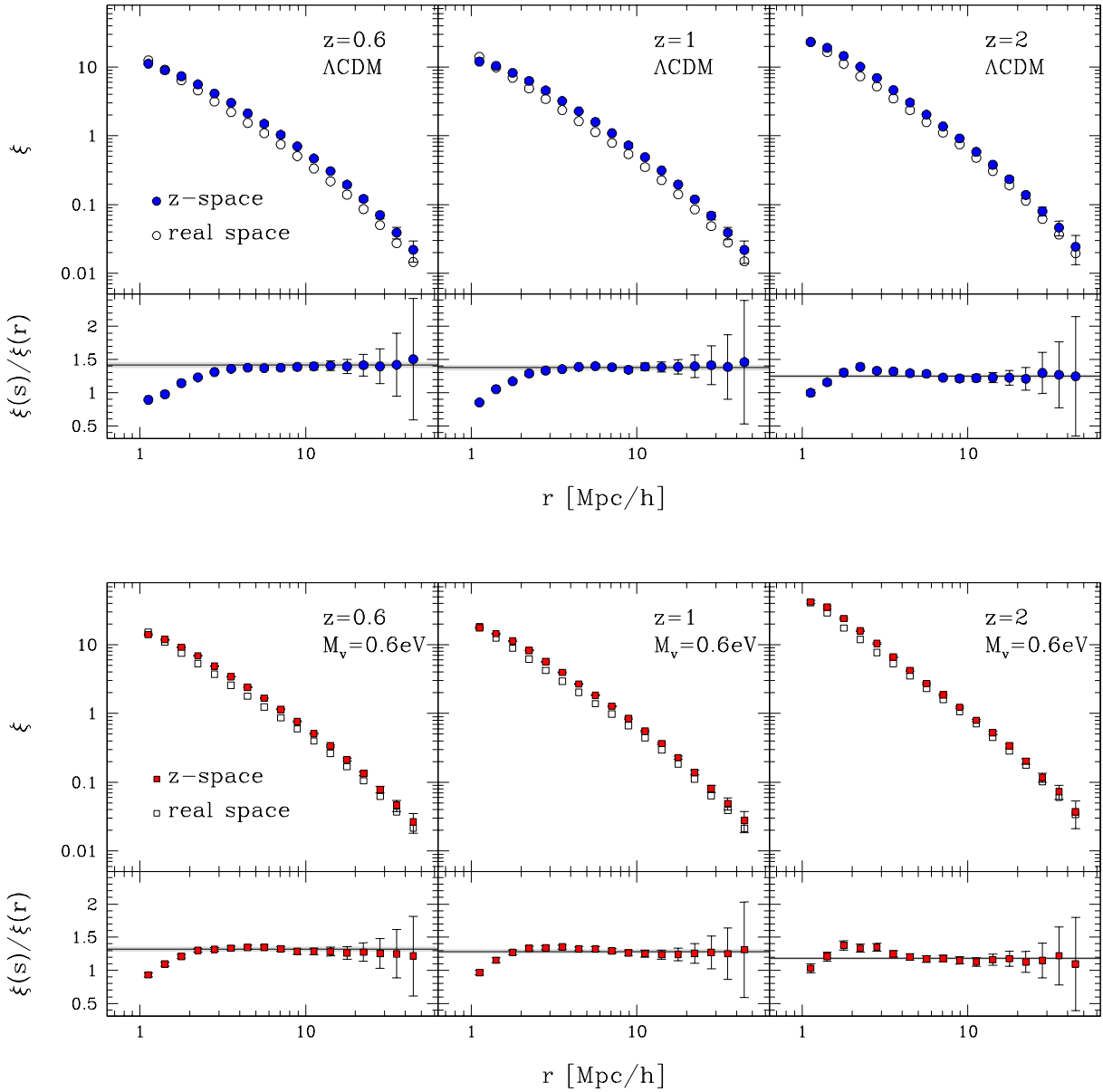}
\caption{Two-point auto-correlation function in real and redshift
  space of the DM haloes in the $\Lambda$CDM N-body simulation (blue
  circles) and in the simulation with $M_\nu=0.6$ eV (red
  squares). The bottom panels show the ratio between them, compared
  with the theoretical expectation given by
  Eq.~(\ref{eq:xiratio}). The error bars represent the statistical
  Poisson noise corrected at large scales as prescribed by
  \citet{mo1992}.}
 \label{fig:xiZspace}
\end{figure*}

\begin{figure*}
\includegraphics[width=\textwidth]{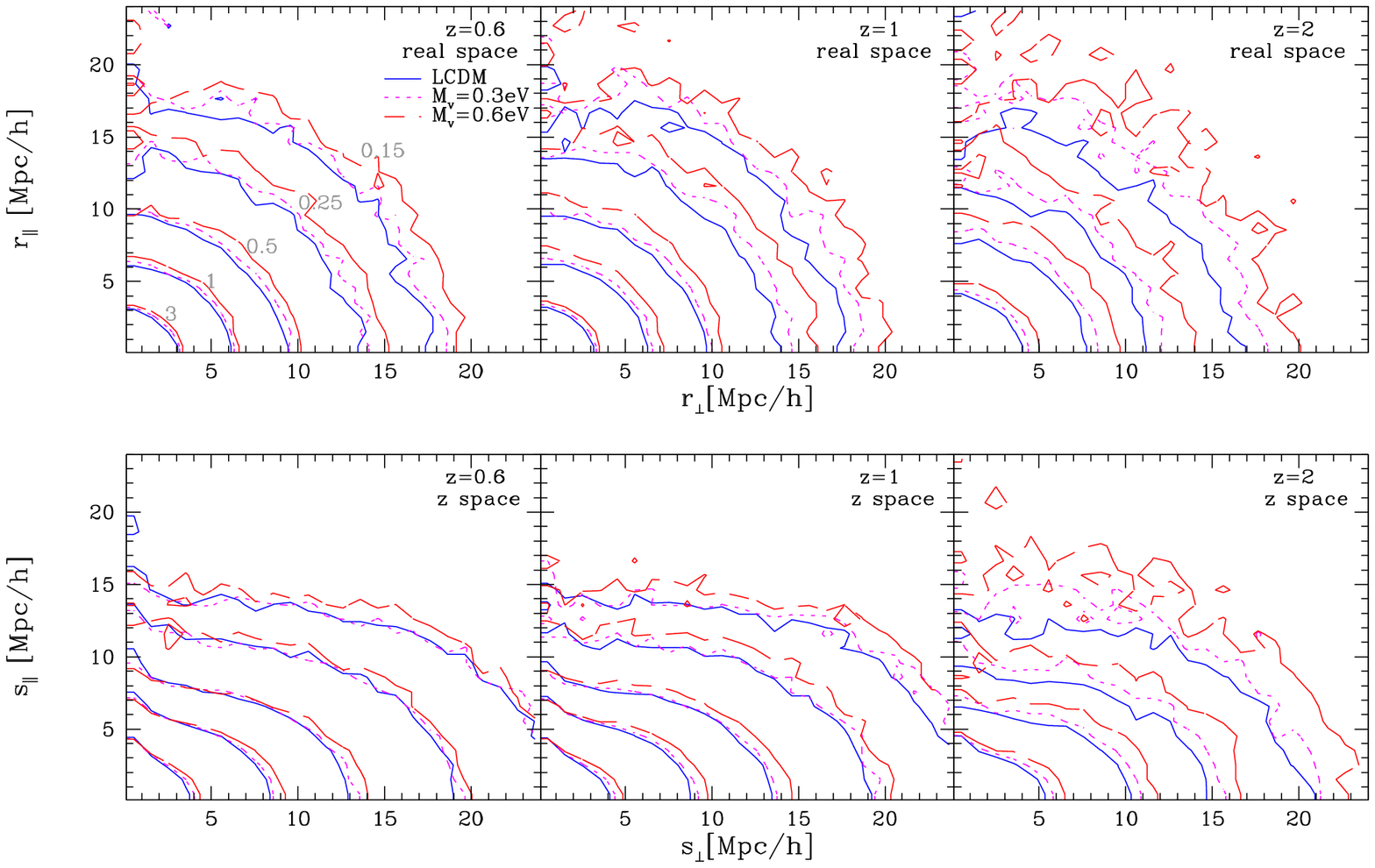}
\caption{Two-point auto-correlation function, $\xi$, in real and
  redshift-space. The contours represent lines of constant
  correlation, \xii=0.15,0.25,0.5,1,3, for $M_\nu=0$ (blue),
  $M_\nu=0.3$ eV (magenta) and $M_\nu=0.6$ (red),
  respectively. Different panels show the
  results at redshifts $z=0.6,1,2$, as labeled.}
 \label{fig:iso}
\end{figure*}

As well known, massive neutrinos strongly affect also the spatial
clustering of cosmic structures. As explained in \S \ref{modelling}, a
standard statistics generally used to quantify the clustering degree of
a population of sources is the two-point auto-correlation
function. Although the free-streaming of massive neutrinos causes a
suppression of the matter power spectrum on scales $k$ larger than the
neutrino free-streaming scale $k_{\rm fs}$, the halo bias results to
be significantly enhanced.  This effect can be physically explained
thinking that, starting from the same $\Delta^2_{\cal R}(k_0)$
as initial condition, due to the suppression of massive neutrino
perturbations,
the same halo bias would correspond, in a $\Lambda$CDM
cosmology without neutrinos, to more massive haloes, which, as well
known, are typically more clustered.

In fact, Fig.~\ref{fig:xiReal} shows, at different redshifts, the
two-point DM halo correlation function measured using the
\citet{landy1993} estimator, compared to the correlation function of
the matter density perturbations.  We observe that, while for a fixed
$\Delta^2_{\cal R}(k_0)$, due to neutrino free-streaming, the total
matter correlation function decreases with respect to the $\Lambda$CDM
case, especially on small scales (compare the solid-blue and
dashed-red lines in Fig.~\ref{fig:xiReal}), the halo correlation
function undergoes the opposite trend (compare the data points in
Fig.~\ref{fig:xiReal}), so that
the matter perturbation suppression is in some way compensated by a
stronger spatial clustering of the massive haloes.

In particular, the halo clustering difference between the $\Lambda$CDM
and $\Lambda$CDM+$\nu$ cosmologies increases with the redshift (as it
happens also for the halo MFs). For $M_\nu=0.6$ eV we find that the
halo correlation function in the presence of massive neutrinos at
$z=1$ is $\sim 20\%$ larger than in a pure $\Lambda$CDM model, and at
$z=2$ the difference rises up to $\sim 40\%$ (see the bottom panels
of Fig.~\ref{fig:xiReal}).

This effect is even more evident in Fig.~\ref{fig:bias1} and
\ref{fig:bias2}, that show the effective bias measured from the
simulations (symbols) compared to the analytical predictions (dotted
lines), obtained using the \citet{sheth_mo_tormen2001} (SMT) bias,
weighted with the ST MF of Eq.~(\ref{eq:MF}):
\begin{equation}
b(z) = \frac{\int_{M_{\rm min}}^{M_{\rm max}} n(M,z) b_{\rm
    SMT}(M,z)dM}{\int_{M_{\rm min}}^{M_{\rm max}} n(M,z)dM},
\label{eq:bias} 
\end{equation}
where $M_{\rm min}$ and $M_{\rm max}$ have been defined in \S
\ref{N-body}.  Also in this case, the theoretical expectations
reproduce correctly the numerical findings, inside the statistical
errors, and, as in the $\Lambda$CDM cosmology, the halo bias results
to be scale-independent on large scales, while the effect of
non-linearities starts to be important for separations $r<20$ \Mpch.

\subsection{Redshift-space distortions}
\label{beta}

As it happens for the halo MFs and clustering, also RSD are strongly
affected by free-streaming neutrinos. Fig.~\ref{fig:xiZspace} shows
the real and redshift-space correlation functions of DM haloes
extracted from the simulations as a function of the neutrino mass. In
the presence of massive neutrinos the rms of galaxy peculiar
velocities is smaller than in a pure $\Lambda$CDM cosmology, due to
the suppression of both the growth rate $f(k,z)$ and the matter power
spectrum $P(k,z)$, which enter the bulk flow predicted by linear theory
\citep{kiakotou2008,elglahav2005}:
\begin{equation}
\langle v^2(R_*) \rangle = (2 \pi^2)^{-1}\; H_0^2 \; \int dk f^2
P_{\rm lin}(k) W_G^2(kR_*)\,,
\end{equation} 
where $W_G(kR_*)$ is the window function, e.g., for a Gaussian sphere
of radius $R_*$, $W(kR_*) \equiv \exp(-k^2 R_*^2/2)$.  This effect
competes with the increase of the halo bias discussed in \S
\ref{clustering}, resulting in a redshift-space halo correlation
function slightly suppressed in a $\Lambda$CDM$+\nu$ cosmology.  In
the bottom panels of Fig.~\ref{fig:xiZspace} we show the ratios
$\xi(s)/\xi(r)$ compared to the theoretical values represented by the
large-scale limit of Eq.~(\ref{eq:ximodellin})
\begin{equation}
\frac{\xi(s)}{\xi(r)} = 1 + \frac{2\beta}{3} + \frac{\beta^2}{5}.
\label{eq:xiratio}
\end{equation}

\begin{figure*}
\includegraphics[width=\textwidth]{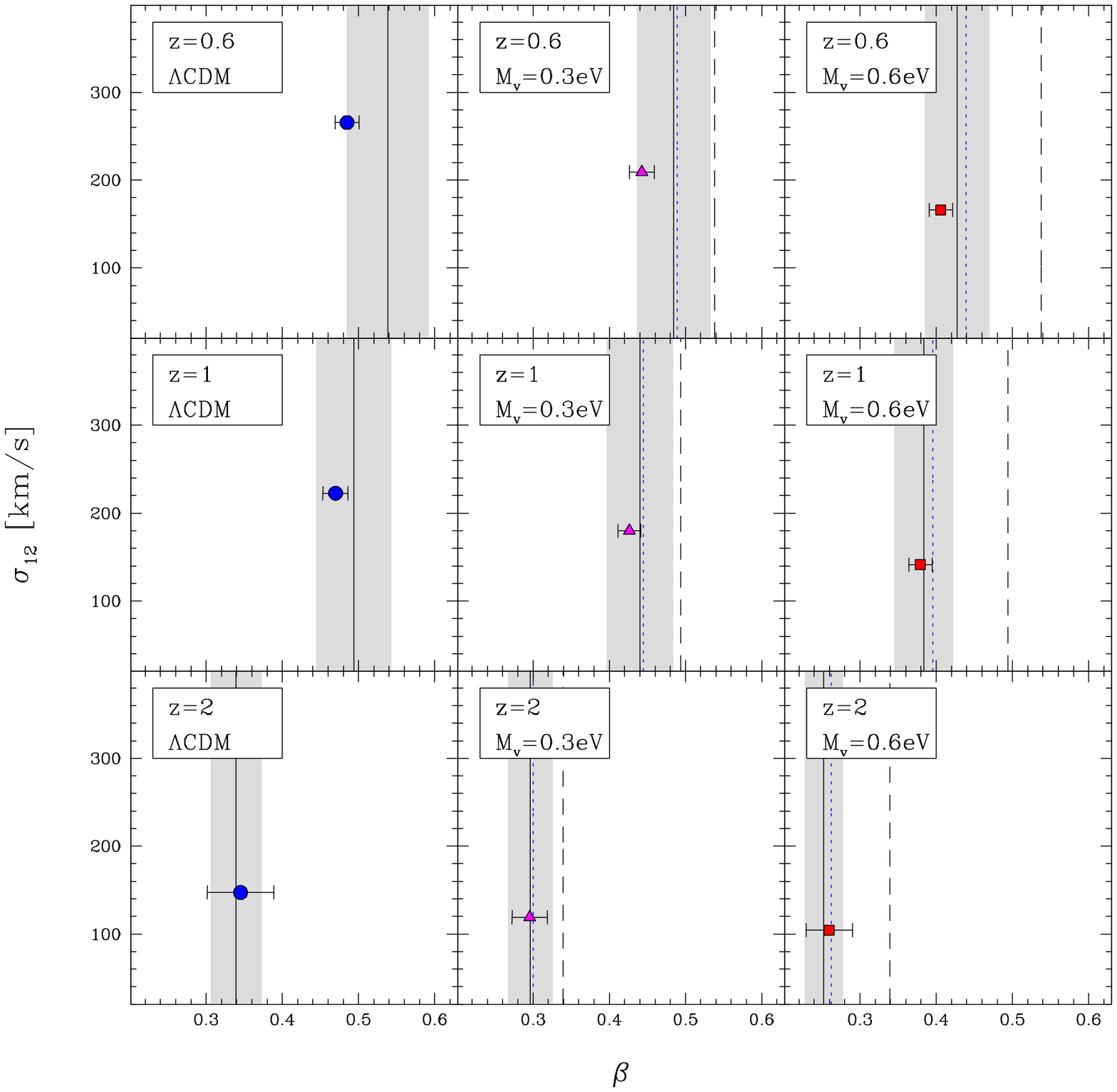}
\caption{Best-fit values of $\beta$-$\sigma_{12}$, as a function of
  $M_\nu$ and redshift (points), compared with the ST theoretical
  predictions (solid lines). The dashed lines show the theoretical
  predictions in a $\Lambda$CDM cosmology with the same
  $\Delta^2_{\cal R}(k_0)$. The blue dotted lines show instead the
  theoretical predictions for a $\Lambda$CDM cosmology normalised to
  the $\sigma_8$ value of the simulation with a massive neutrino
  component, as explained in the text. The shaded grey bands represent
  the propagated $\sim 10\%$ theoretical bias error. The error bars
  represent the scatter in the measured $\beta$ obtained dividing the
  simulation box in 27 sub-boxes, and rescaled by the square root of
  the total volume of the simulation box \citep{guzzo2008}.}
 \label{fig:beta}
\end{figure*}

The effect of massive neutrinos on RSD is evident in particular when
the correlation function is measured as a function of the two
directions perpendicular and parallel to the line-of-sight. In fact,
from the top and bottom panels of Fig.~\ref{fig:iso}, we observe that,
in the case of massive neutrinos, the spatial halo clustering is less
enhanced in redshift-space than in real-space. On large scales, this
effect is due to the lower value of $\langle v^2(R_*) \rangle$ when
neutrinos free-streaming is taken into account.  On small scales, our
analysis shows that also FoG get decreased in the presence of massive
neutrinos, so that the best-fit values of $\beta$ and $\sigma_{\rm
  12}$, derived by modelling galaxy clustering anisotropies, result to
be different than what expected in a $\Lambda$CDM cosmology.  This
might induce a bias in the inferred growth rate from data analysis,
and therefore a potentially false signature of modified gravity
\citep[see e.g.][]{simpson2011}.  Moreover, estimates of $\beta$ and
$\sigma_{\rm 12}$, when compared with the $\Lambda$CDM expectations,
yield an indirect neutrino mass measurement and may help breaking
degeneracies with the other cosmological parameters.

We quantify these effects in Fig.~\ref{fig:beta}, which shows the
best-fit values of $\beta$ and $\sigma_{12}$ as a function of $M_\nu$
and $z$, where we have neglected their scale-dependence which, for the
neutrino masses considered in this work, is small enough that
statistical errors hide deviations of $\beta$ and $\sigma_{12}$ from
spatial uniformity \citep{kiakotou2008}.  Therefore, in this case we
have considered the linear redshift-space distortion parameter as a
function of the redshift alone, $\beta =f(z)/b(z)\simeq
\Omega_m(z)^\gamma/b(z)$, with $\gamma=0.545$. The data points of
Fig.~\ref{fig:beta} show that neutrinos free-streaming suppresses
$\beta$ and $\sigma_{12}$ by an amount which increases with $M_\nu$
and $z$, and, fixed $\Delta^2_{\cal R}(k_0)$, is clearly
distinguishable from the corresponding $\Lambda$CDM values (dashed
lines).  As an example, at $z=0.6$ the $\beta$ best-fit values
decrease by $\sim 10\%$ for $M_\nu=0.3$ eV, and by $\sim 25\%$ for
$M_\nu=0.6$ eV. Likewise, the $\sigma_{12}$ best-fit values decrease
by $\sim 25\%$ for $M_\nu=0.3$ eV, and by $\sim 45\%$ for $M_\nu=0.6$.

On the other hand, the $\beta$ best-fit values fall in the shaded grey
bands, which represent the propagated $\sim 10\%$ theoretical bias
error. These bands contain also the theoretical predictions obtained
in a $\Lambda$CDM cosmology renormalised with the $\sigma_8$ value of
the simulations with a massive neutrino component (blue dotted
lines). This means that, if an error of $\sim 10\%$ is assumed on bias
measurements, we are not able to distinguish the effect of massive
neutrinos on $\beta$ when the two cosmological models with and without
$\nu$ are normalised to the same $\sigma_8$.

In Fig.~\ref{fig:beta_error} we show, as a function of $M_\nu$ and
$z$, the relative difference between the theoretical $\beta$ values
calculated in the $\Lambda$CDM+$\nu$ and $\Lambda$CDM cosmologies,
normalised to the same $\sigma_8$.  At $z=1$ and for $M_\nu>0.6$ eV,
the relative difference with respect to the $M_\nu=0$ case is
$\Delta\beta/\beta \gtrsim 3\%$. This result is interesting, since
future spectroscopic galaxy surveys, as EUCLID, JEDI and WFIRST,
should be able to measure the linear redshift-space distortion
parameter with errors $\leq 3\%$ at $z\leq 1$, per redshift bin.

\section{Conclusions} 
\label{conclu}
In this work we have studied the effect of cosmological neutrinos on
the DM halo mass function, clustering properties and redshift-space
distortions.  To this purpose we have exploited the grid
implementation of the hydrodynamical N-body simulations developed by
\citet{viel2010}, which include a massive neutrino component, taking
into account the effect of neutrinos free-streaming on the cosmic
structure evolution.  In order to model RSD, we have adopted the
so-called streaming model \citep{peebles1980}, which consists of
linear theory and a convolution on the line-of-sight with a velocity
distribution. This model is accurate enough to robustly constrain the
effect of massive neutrinos on RSD when applied on scales $\lesssim
50$ \Mpch.

We have compared the findings from the $\Lambda$CDM and the
$\Lambda$CDM+$\nu$ simulations, and analysed their agreement with the
analytical predictions of ST \citep{sheth_mo_tormen2001,sheth2002}.
Concerning the halo MF, we recover what theoretically expected,
i.e. that, starting from the same $\Delta^2_{\cal R}(k_0)$ as initial
condition, massive neutrinos suppress the comoving number density of
DM haloes by an amount that increases with the total neutrino mass
$M_\nu$. The suppression affects mainly haloes of mass $10^{14}
M_\odot/h <M<10^{15} M_\odot/h$, depending slightly on the redshift
$z$.  As an example, the number density of haloes with mass $10^{14}
M_\odot/h$ at $z=0$ decreases by $\sim 15\%$ for $M_\nu=0.3$ eV and by
$\sim 30\%$ for $M_\nu=0.6$ eV, and, at $z=1$, by$\sim 40\%$ and $\sim
70\%$, respectively. Moreover, with increasing $z$, the suppression of
the halo number density due to free-streaming neutrinos moves towards
masses $M \leq 10^{14} M_\odot/h$.

With regard to the halo clustering in the real-space, we observe that
the trend of the halo correlation function $\xi(r)$ is opposite to the
dark matter one.  In fact, on one side, for a fixed $\Delta^2_{\cal
  R}(k_0)$ the total matter correlation function decreases with
respect to the $\Lambda$CDM case due to neutrino free-streaming, in
particular on small scales. On the other side, the halo correlation
function undergoes the opposite trend since the halo bias results to
be significantly enhanced. For $M_\nu=0.6$ eV, we find that the halo
correlation function in the presence of massive neutrinos at $z=1$ is
$\sim 20\%$ larger than in a pure $\Lambda$CDM model, and at $z=2$
this difference rises up to $\sim 40\%$.  Also in this case, the
theoretical ST bias model reproduces correctly the numerical findings,
inside the statistical errors, and, as in the $\Lambda$CDM cosmology,
the halo bias results to be scale-independent on scales larger than $r
\gtrsim 20$ \Mpch.

Considering RSD, we find that the rise of the spatial halo clustering
due to massive neutrinos is less enhanced in the redshift-space than
in the real-space.  In fact, on large scales, the value assumed by the
bulk flow, $\langle v^2(R_*) \rangle$, in a $\Lambda$CDM+$\nu$ cosmology
is smaller than in a pure $\Lambda$CDM one.  On small scales, also FoG
get decreased in the presence of massive neutrinos, so that the
best-fit values of $\beta$ and $\sigma_{12}$ reduce by an amount which
increases with $M_\nu$ and $z$. As an example, fixed the same initial
condition on $\Delta^2_{\cal R}(k_0)$, at $z=0.6$ the $\beta$ best-fit
values decrease by $\sim 10\%$ for $M_\nu=0.3$ eV, and by $\sim 25\%$
for $M_\nu=0.6$ eV. Likewise, the $\sigma_{12}$ best-fit values
decrease by $\sim 25\%$ for $M_\nu=0.3$ eV, and by $\sim 45\%$ for
$M_\nu=0.6$.

\begin{figure}
\includegraphics[width=0.43\textwidth]{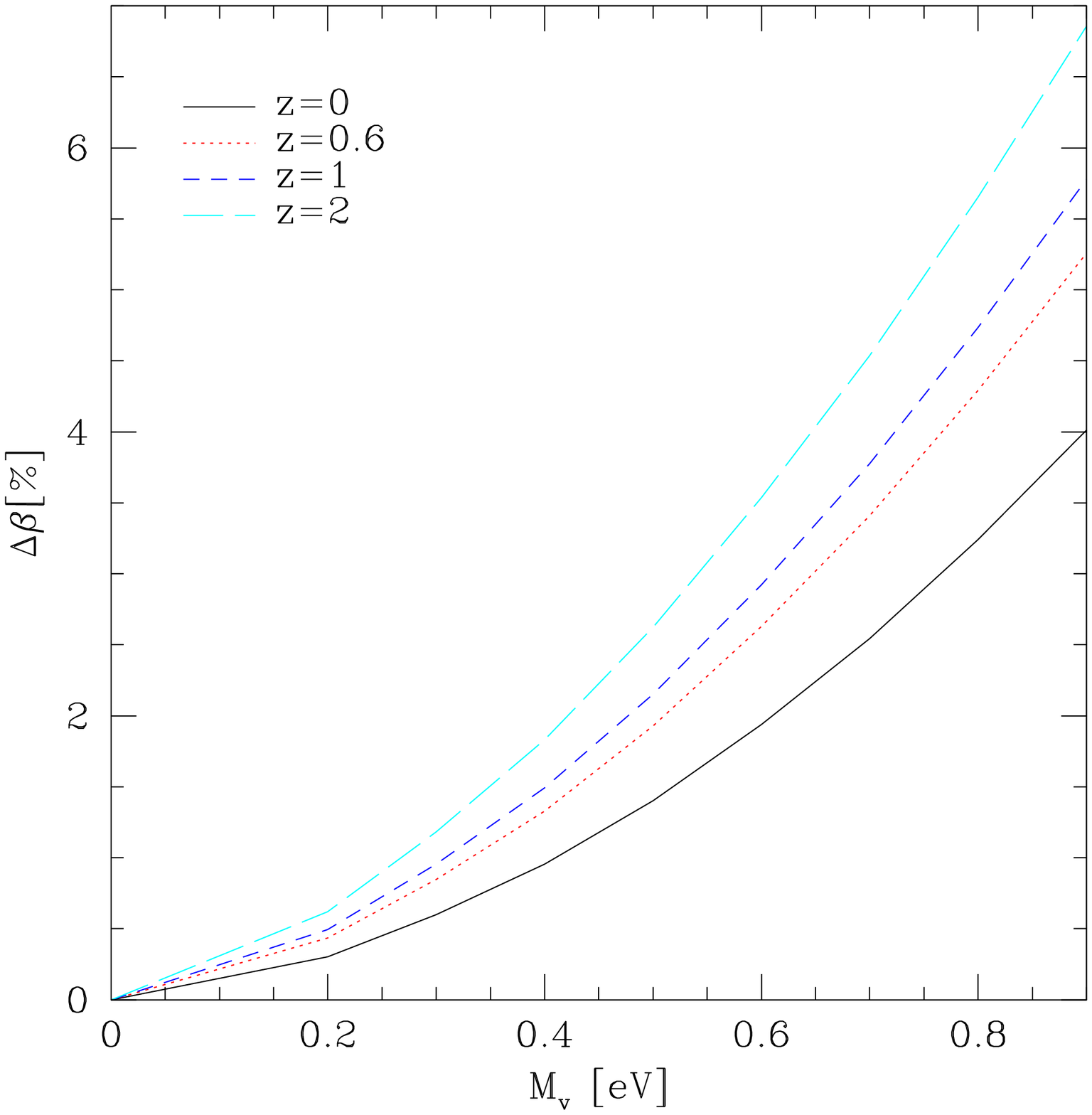}
\caption{The relative difference between the theoretical $\beta$
  values calculated in the $\Lambda$CDM+$\nu$ and $\Lambda$CDM
  cosmologies, normalised to the same $\sigma_8$.}
\label{fig:beta_error}
\end{figure}

If not taken correctly into account, these effects could lead to a
potentially fake signatures of modified gravity.  Moreover, estimates
of $\beta$ and $\sigma_{\rm 12}$ can be used to extract measurements
of the total neutrino mass and may help breaking degeneracies with the
other cosmological parameters.

However, these effects are nearly perfectly degenerate with the
overall amplitude of the matter power spectrum as characterised by
$\sigma_8$. This strong $M_\nu$-$\sigma_8$ degeneracy undermines the
potentiality of the mentioned methods in constraining the neutrino
mass. For instance, the difference between the halo MFs in the
$\Lambda$CDM+$\nu$ and $\Lambda$CDM models largely decreases if we
normalise the two cosmologies to the same $\sigma_8$. Similarly, when
analysing RSD, we find that the $\beta$ best-fit values fall in the
shaded grey bands of Fig.~\ref{fig:beta}, representing the propagated
$\sim 10\%$ theoretical bias error, and which contain the theoretical
predictions obtained in a $\Lambda$CDM cosmology renormalised with the
$\sigma_8$ value of the simulations with a massive neutrino
component. For such a value of the bias error, we are prevented to
distinguish the effect of massive neutrinos on $\beta$, if we use as
initial condition the same $\sigma_8$ value both for the
$\Lambda$CDM+$\nu$ and $\Lambda$CDM cosmologies.

Nonetheless, the $\sigma_8$ renormalisation of the matter power
spectrum does not totally cancel the neutrino effects which, in this
case, depending on the $M_\nu$ value, alter the MF shape and amplitude
especially for the less massive objects. As an example, at $z=0$ the
difference between the halo MFs with and without massive neutrinos is
$\sim 3\%$ at $M=10^{13}M_\odot/h$ for $M_\nu=0.6$ eV. The detection
of this small effect depends on the sensitivity in measuring the halo
MF at masses $M<10^{14}M_\odot/h$.

More promising are measurements of $\beta$. In
Fig.~\ref{fig:beta_error} we show, as a function of $M_\nu$ and $z$,
the relative difference between the theoretical $\beta$ values
calculated in the $\Lambda$CDM+$\nu$ and $\Lambda$CDM cosmologies,
normalised to the same $\sigma_8$.  At $z=1$ and for $M_\nu>0.6$ eV,
the relative difference with respect to the $M_\nu=0$ case is
$\Delta\beta/\beta \gtrsim 3\%$. This results is interesting, since
future nearly all-sky spectroscopic galaxy surveys, like EUCLID, JEDI
and WFIRST, should be able to measure the linear redshift-space
distortion parameter with errors $\lesssim 3\%$ at $z\leq 1$, per
redshift bin. This means that, even exploiting information from
$\beta$ measurements alone, they will contribute, along with other
cosmological probes, to constrain the value of the total mass of
cosmological neutrinos.

\section*{acknowledgments}
We warmly thank K. Dolag, E. Branchini, M. Haehnelt and V. Springel
for precious suggestions. We would also like to thank the anonymous
referee for helpful comments. We acknowledge financial contributions
from contracts ASI-INAF I/023/05/0, ASI-INAF I/088/06/0, ASI
I/016/07/0 ’COFIS’, ASI ’Euclid-DUNE’ I/064/08/0, ASI-Uni
Bologna-Astronomy Dept. ’Euclid-NIS’ I/039/10/0, and PRIN MIUR ’Dark
energy and cosmology with large galaxy surveys’. MV is supported by a
PRIN-INAF 2009 "Towards an italian network for computational
cosmology", ASI/AAE, INFN/PD-51, PRIN-MIUR 2008 and the FP7 ERC
Starting Grant "cosmoIGM".  Simulations were performed at the HPCS
Darwin Supercomputer center and post-processed at CINECA.

\bibliographystyle{mn2e} \bibliography{/home/frea/papers/bib}

\end{document}